\documentclass[number,preprint,3p]{elsarticle}

\usepackage{color}
\usepackage{graphicx}
\usepackage{subcaption}
\usepackage{algorithm}
\usepackage{bm}
\usepackage[colorlinks]{hyperref}
\usepackage{amssymb}
\usepackage{amsthm}
\usepackage{amsmath}
\usepackage{amssymb}
\usepackage{mathtools}
\usepackage{dsfont}
\usepackage{booktabs}
\usepackage{url}
\usepackage{scrextend}
\usepackage{epstopdf}
\usepackage{float}
\usepackage{tcolorbox}
\usepackage{tablefootnote}
\usepackage[perpage]{footmisc}
\usepackage{bbm}
\usepackage{lineno}

\def\rn{\mathbb{R}^n}

\newcommand{\vect}[1]{\boldsymbol{#1}}


\biboptions{numbers,sort&compress,square}
\journal{arXiv}

\begin{document}
		\begin{frontmatter}
			\renewcommand{\thefootnote}{\fnsymbol{footnote}}
			\title{Relaxation-based importance sampling for structural reliability analysis}
			\author[1]{Jianhua Xian}
			\author[1]{Ziqi Wang\footnotemark[1]}
			\address[1]{Department of Civil and Environmental Engineering, University of California, Berkeley, United States}
			\footnotetext[1]{Corresponding author. \href{mailto:ziqiwang@berkeley.edu}{ziqiwang@berkeley.edu}}
			\begin{abstract}
					This study presents an importance sampling formulation based on adaptively relaxing parameters from the indicator function and/or the probability density function. The formulation embodies the prevalent mathematical concept of relaxing a complex problem into a sequence of progressively easier sub-problems. Due to the flexibility in constructing relaxation parameters, relaxation-based importance sampling provides a unified framework for various existing variance reduction techniques, such as subset simulation, sequential importance sampling, and annealed importance sampling. More crucially, the framework lays the foundation for creating new importance sampling strategies, tailoring to specific applications. To demonstrate this potential, two importance sampling strategies are proposed. The first strategy couples annealed importance sampling with subset simulation, focusing on low-dimensional problems. The second strategy aims to solve high-dimensional problems by leveraging spherical sampling and scaling techniques. Both methods are desirable for fragility analysis in performance-based engineering, as they can produce the entire fragility surface in a single run of the sampling algorithm. Three numerical examples, including a 1000-dimensional stochastic dynamic problem, are studied to demonstrate the proposed methods.
			\end{abstract}
			
			\begin{keyword}
				importance sampling \sep fragility analysis \sep reliability analysis 
				
			\end{keyword}
			
		\end{frontmatter}
		
		\renewcommand{\thefootnote}{\fnsymbol{footnote}}
		
		\section{Introduction}
		
		\noindent The computational challenges of reliability analysis are featured by expensive computational models, high-dimensional uncertainties, and low failure probabilities. An increasing amount of reliability methods are developed to address these challenges, e.g., the first- and second-order reliability methods \cite{ditlevsen1996structural,der2005first}, moment-based methods \cite{zhao2001moment,dang2020mixture,xu2022adaptive,xi2012comparative}, surrogate modeling-based methods \cite{rajashekhar1993new,echard2011ak,wang2020novel,dang2022parallel,dhulipala2022reliability,dhulipala2022active}, and probability density evolution-based methods \cite{li2009stochastic,chen2019direct,xian2021seismic,lyu2022unified}. These reliability methods more or less suffer from the curse of dimensionality, restriction to specialized problems, and inaccuracy for low-probability events. In comparison, sampling methods \cite{rubinstein1998modern} are typically general in applicability, insensitive to the dimensionality and nonlinearity of the computational model, and have the theoretical guarantee to converge to the correct solution. Therefore, although computationally more demanding than approximate methods such as the first-order reliability method, sampling methods are a promising research direction that complements the research effort of other reliability methods.  
		
		Among various sampling methods, the direct Monte Carlo is general in applicability and insensitive to the curse of dimensionality; however, it has a slow convergence rate of $\mathcal{O}(N^{-1/2})$, where $N$ is the number of random samples. The slow convergence rate leads to a prohibitively computational cost in rare event simulation, which is a key component in the risk, reliability, and resilience assessment of structures subjected to natural hazards. This motivates the study of various variance reduction techniques. The importance sampling \cite{fujita1988updating,melchers1989importance,melchers2001estimation,rubinstein2004cross,de2005tutorial,chan2011comparison,kurtz2013cross,wang2016cross,yang2017cross,geyer2019cross} accelerates probability estimations by sampling from an importance density. The efficiency of importance sampling hinges on selecting a proper importance density: the adopted density should resemble the theoretical optimal importance density. A classic importance sampling strategy is to construct a multivariate Gaussian density centered at the design point \cite{fujita1988updating,melchers1989importance,melchers2001estimation}. This strategy has weak scalability for high-dimensional problems; moreover, it becomes inefficient with numerous failure modes.  A more advanced scheme is to optimize a parametric importance density model by adaptively tuning the parameters \cite{rubinstein2004cross,de2005tutorial,chan2011comparison}. The parametric density models can be mixture distributions, such as the Gaussian mixture and von Mises-Fisher mixture \cite{kurtz2013cross,wang2016cross,yang2017cross,geyer2019cross}, so that the importance density can capture multiple failure modes. However, in general, a parametric importance density model may not work well for complex reliability problems with high-dimensional uncertainties. The multifidelity importance sampling \cite{peherstorfer2018survey,peherstorfer2016multifidelity,pham2022ensemble} is a promising approach to address the limitations of parametric importance density models. It leverages a low-fidelity surrogate model to construct the importance density to accelerate the sampling associated with the original high-fidelity computational model. Recently, the concept of multifidelity uncertainty quantification has been introduced into the dynamic structural reliability analysis in the context of equivalent linearization for random vibration problems, in which an equivalent linear model is optimized as a low-fidelity
		model to accelerate the sampling of the original high-fidelity model \cite{wang2022optimized}. The readers can refer to \cite{tabandeh2022review} for a comprehensive review of importance sampling methods.
		
		Another way to construct variance reduction techniques is to leverage non-parametric approaches such as Markov Chain Monte Carlo to sequentially estimate the target probability. The subset simulation \cite{au2001estimation,au2005reliability,li2010design,diazdelao2017bayesian,betz2018bayesian}, sequential importance sampling \cite{neal2005estimating,del2006sequential,katafygiotis2007estimation,papaioannou2016sequential,papaioannou2018reliability,beaurepaire2013reliability}, and annealed importance sampling \cite{neal2001annealed,lyman2007annealed,stordal2015iterative} all have a mechanism of sequential approximation, and they emerged around the same period. The subset simulation was first proposed for rare event probability estimation in the reliability community \cite{au2001estimation}; later, it was further developed for the reliability-based design optimization \cite{au2005reliability,li2010design} and Bayesian analysis \cite{diazdelao2017bayesian,betz2018bayesian}. The basic
		idea of subset simulation is to decompose the target failure event into a series of nested intermediate failure events, and the failure probability is expressed as a product of conditional probabilities of  intermediate failure events. The sequential importance sampling (also known as bridge or linked importance sampling) was initially proposed in the statistical community for estimating normalizing constants in Bayesian analysis \cite{neal2005estimating,del2006sequential}, and later it was applied to structural reliability analysis \cite{katafygiotis2007estimation,papaioannou2016sequential}, reliability sensitivity analysis \cite{papaioannou2018reliability}, and reliability-based optimization \cite{beaurepaire2013reliability}. For structural reliability analysis, the sequential importance sampling constructs a series of smooth intermediate distributions to gradually approximate the optimal importance density, and the failure probability is evaluated by adaptive sampling from these intermediate distributions. The annealed importance sampling was also initially proposed in the statistical community \cite{neal2001annealed,lyman2007annealed,stordal2015iterative}, but to our knowledge, the annealed importance sampling has never been investigated in the context of structural reliability analysis. In annealed importance sampling, the intermediate distributions take the exponential forms of the target
		distribution, and tuning the exponent has a physical analogy of an annealing process. It is noted that, in the above sequential sampling methods, a major computational challenge is the sampling from complex, implicitly defined intermediate distributions, which requires the use of advanced Markov Chain Monte Carlo algorithms \cite{miao2011modified,papaioannou2015mcmc,neal2011mcmc,wang2019hamiltonian}.
		
		In fact, subset simulation, sequential importance sampling, and annealed importance sampling all embody a prevalent mathematical concept of relaxing a complex problem into a sequence of progressively easier sub-problems. This study presents an importance sampling formulation based on adaptively relaxing parameters from the indicator function and/or the probability density function. The proposed relaxation-based importance sampling offers a broad framework for developing various importance sampling strategies. Subset simulation, sequential importance sampling, and annealed importance sampling belong to the relaxation-based importance sampling framework with a single relaxation parameter. More critically, the framework creates the possibility for developing new variants of importance sampling, adapting to the needs of different problems. To demonstrate the potential of the proposed framework, we develop two new importance sampling strategies. The first strategy couples annealed importance sampling with subset simulation, while the second leverages spherical sampling and scaling techniques \cite{wang2016cross,wang2018hyper}. Both methods are desirable for fragility analysis in performance-based engineering \cite{porter2007creating,gunay2013peer}, as they can produce the entire fragility surface in a single run of the sampling algorithm.
		
		The paper introduces the general formulation of relaxation-based importance sampling in Section \ref{Framework}. Section \ref{Newvariant}  develops two importance sampling strategies in the framework of relaxation-based importance sampling, aiming to efficiently solve fragility problems in performance-based engineering. The numerical examples in Section \ref{Application} demonstrate the performance of the proposed methods for low- and high-dimensional reliability/fragility problems. Section \ref{conclude} provides concluding remarks. To optimize readability, some important but highly technical implementation details are introduced in \ref{App: Extrapolation} and \ref{Append:SolutionProcedureAIS}. 
		
		\section{General formulation of the relaxation-based importance sampling}\label{Framework}
		\noindent For a reliability problem defined in the $n$-dimensional uncorrelated standard normal space, the failure probability can be formulated as
		\begin{equation}\label{FirstEq}
			P_{f}=\int _{\vect{x}\in \rn}\mathbb{I}(G(\vect{x})\leqslant 0)f_{\vect{X}}(\vect{x})\mathrm{d}\vect{x}\,,
		\end{equation}
		where $f_{\vect{X}}(\vect{x})$ is the joint probability density function (PDF) for $n$ independent standard Gaussian variables, $G(\vect{x})$ is the limit-state function, and $\mathbb{I}(G(\vect{x})\leqslant 0)$ is the binary indicator function that gives 1 if $G(\vect{x})\leqslant 0$ and 0 otherwise.
		
		The main idea of importance sampling (IS) is to introduce an importance density, denoted as $h(\vect{x})$, to Eq.\eqref{FirstEq}:
		\begin{equation}\label{SecondEq}
			P_{f}=\int _{\vect{x}\in \rn}\mathbb{I}(G(\vect{x})\leqslant 0)\frac{f_{\vect{X}}(\vect{x})}{h(\vect{x})}h(\vect{x})\mathrm{d}\vect{x}\,.
		\end{equation}
		A valid importance density should dominate $\mathbb{I}(G(\vect{x})\leqslant0)f_{\vect{X}}(\vect{x})$, i.e., $h(\vect{x})=0\Rightarrow\mathbb{I}(G(\vect{x})\leqslant0)f_{\vect{X}}(\vect{x})=0$; this condition suggests that the failure domain should be covered by the support of the importance density. The theoretical optimal importance density has the following form \cite{rubinstein1998modern}:
		\begin{equation}\label{ThirdEq}
			h^{\ast }(\vect{x})=\frac{\mathbb{I}(G(\vect{x})\leqslant 0)f_{\vect{X}}(\vect{x})}{P_{f}}\,.
		\end{equation}
		
		It is straightforward to verify that the optimal importance density leads to zero-variance failure probability estimation. However, the normalizing constant (denominator) of the optimal density is the (to be solved) failure probability. This “Catch-22” situation does have a practical way out: an importance density model can be constructed and optimized to imitate the numerator of the optimal importance density.
		
		The core concept of the proposed relaxation-based importance sampling (RIS) is to construct relaxation parameters such that by adaptively tuning them, a rare-event probability can be estimated through computations of a sequence of larger probabilities. The relaxation parameters can be applied to the indicator function and/or the probability density function. Formally, we define $\eta_{T}(\vect{x}):=\mathbb{I}(G(\vect{x})\leqslant 0)f_{\vect{X}}(\vect{x})$, where $T\in\mathbb{N}^+$ indexes the final stage of the tuning sequence. Next, we design relaxation parameters $\vect{\lambda }_T$ and introduce them into $\eta_{T}(\vect{x})$, resulting in $\eta_{T}(\vect{x};\vect{\lambda } _{T})$. The values of $\vect{\lambda } _{T}$ are specified such that $\eta_{T}(\vect{x};\vect{\lambda } _{T})$ is equivalent to $\mathbb{I}(G(\vect{x})\leqslant 0)f_{\vect{X}}(\vect{x})$. For example, subset
		simulation introduces a $\lambda  _{T}$ into the indicator function, resulting in $\eta_{T}(\vect{x};\lambda _{T})=\mathbb{I}(G(\vect{x})\leqslant\lambda _{T})f_{\vect{X}}(\vect{x})$, where $\lambda_{T}=0$. The RIS considers a more general construction, and thus at the current stage, we do not explicitly specify how/where to introduce the relaxation parameters. Provided with $\vect{\lambda } _{T}$, we tune the values of $\vect{\lambda } _{T}$ to $\vect{\lambda }_{1}$ so that the probability/integral associated with $\eta_{1}(\vect{x};\vect{\lambda } _{1})$ is sufficiently large. In other words, a relaxed failure event specified by $\eta_{1}(\vect{x};\vect{\lambda } _{1})$ is constructed with the probability
		\begin{equation}\label{fourthEq}
			P_{1}=\int _{\vect{x}\in \rn}\eta_{1}(\vect{x};\vect{\lambda } _{1})\mathrm{d}\vect{x}\,.
		\end{equation}
		Given that $P_{1}$ is estimated, the optimal importance density with respect to the relaxed failure event has a form similar to Eq.\eqref{ThirdEq}:
		\begin{equation}\label{fifthEq}
			h_{1}^{\ast }(\vect{x};\vect{\lambda } _{1})=\frac{\eta_{1}(\vect{x};\vect{\lambda } _{1})}{P_{1}}\,.
		\end{equation}
		Using the optimal importance density $h_{1}^{\ast}(\vect{x};\vect{\lambda} _{1})$, one can estimate the probability of a less relaxed failure event specified by $\eta_{2}(\vect{x};\vect{\lambda} _{2})$, i.e.,
		\begin{equation}\label{sixthEq}
			P_{2}=\int _{\vect{x}\in \rn}\frac{\eta_{2}(\vect{x};\vect{\lambda} _{2})}{h_{1}^{\ast}(\vect{x};\vect{\lambda} _{1})}h_{1}^{\ast}(\vect{x};\vect{\lambda} _{1})\mathrm{d}\vect{x}=P_{1}\int _{\vect{x}\in \rn}\frac{\eta_{2}(\vect{x};\vect{\lambda} _{2})}{\eta_{1}(\vect{x};\vect{\lambda} _{1})}h_{1}^{\ast}(\vect{x};\vect{\lambda} _{1})\mathrm{d}\vect{x}\,.
		\end{equation}
		Given that $P_{2}$ is estimated, one can construct the optimal importance density corresponding to $\eta_{2}(\vect{x};\vect{\lambda} _{2})$, i.e.,
		\begin{equation}\label{seventhEq}
			h_{2}^{\ast }(\vect{x};\vect{\lambda} _{2})=\frac{\eta_{2}(\vect{x};\vect{\lambda} _{2})}{P_{2}}\,.
		\end{equation}
		Using $h_{2}^{\ast }(\vect{x};\vect{\lambda} _{2})$, one can subsequently estimate the probability associated with $\eta_{3}(\vect{x};\vect{\lambda} _{3})$:
		\begin{equation}\label{eighthEq}
			P_{3}=\int _{\vect{x}\in \rn}\frac{\eta_{3}(\vect{x};\vect{\lambda} _{3})}{h_{2}^{\ast}(\vect{x};\vect{\lambda} _{2})}h_{2}^{\ast}(\vect{x};\vect{\lambda} _{2})\mathrm{d}\vect{x}=P_{2}\int _{\vect{x}\in \rn}\frac{\eta_{3}(\vect{x};\vect{\lambda} _{3})}{\eta_{2}(\vect{x};\vect{\lambda} _{2})}h_{2}^{\ast}(\vect{x};\vect{\lambda} _{2})\mathrm{d}\vect{x}\,.
		\end{equation}
		Proceeding in this manner, the probability of the relaxed failure event specified by $\eta_{j+1}(\vect{x};\vect{\lambda} _{j+1})$ can be obtained as
		\begin{equation}\label{ninethEq}
			P_{j+1}=\int _{\vect{x}\in \rn}\frac{\eta_{j+1}(\vect{x};\vect{\lambda} _{j+1})}{h_{j}^{\ast}(\vect{x};\vect{\lambda} _{j})}h_{j}^{\ast}(\vect{x};\vect{\lambda} _{j})\mathrm{d}\vect{x}=P_{j}\int _{\vect{x}\in \rn}\frac{\eta_{j+1}(\vect{x};\vect{\lambda} _{j+1})}{\eta_{j}(\vect{x};\vect{\lambda} _{j})}h_{j}^{\ast}(\vect{x};\vect{\lambda} _{j})\mathrm{d}\vect{x}\,,
		\end{equation}
		where $h_{j}^{\ast}(\vect{x};\vect{\lambda} _{j})=\frac{\eta_{j}(\vect{x};\vect{\lambda} _{j})}{P_{j}}$ is the optimal importance density corresponding to $\eta_{j}(\vect{x};\vect{\lambda} _{j})$. The above process is repeated until the target failure event specified by $\eta_{T}(\vect{x};\vect{\lambda} _{T})$ is reached. The failure probability can be expressed as
		\begin{equation}\label{tenthEq}
			\begin{split}
				P_{f}&=P_{T-1}\int _{\vect{x}\in \rn}\frac{\eta_{T}(\vect{x};\vect{\lambda} _{T})}{\eta_{T-1}(\vect{x};\vect{\lambda} _{T-1})}h_{T-1}^{\ast}(\vect{x};\vect{\lambda} _{T-1})\mathrm{d}\vect{x}\\&=P_{1}\prod _{j=1}^{T-1}\int _{\vect{x}\in \rn}\frac{\eta_{j+1}(\vect{x};\vect{\lambda} _{j+1})}{\eta_{j}(\vect{x};\vect{\lambda} _{j})}h_{j}^{\ast}(\vect{x};\vect{\lambda} _{j})\mathrm{d}\vect{x}\,.
			\end{split}
		\end{equation}
		
		Eq.\eqref{tenthEq} offers the general formulation of RIS. By constructing different  relaxation parameters, various IS strategies can be developed. \textbf{Most crucially, by properly designing relaxation parameters, the intermediate probabilities from Eq.\eqref{ninethEq}--byproducts of RIS--can serve specific applications}. In the following subsections, we will show that subset simulation \cite{au2001estimation}, sequential importance sampling \cite{papaioannou2016sequential}, and annealed importance sampling \cite{neal2001annealed} all fall into the framework of RIS.
		
		\subsection{Subset simulation}\label{SS}
		\noindent In subset simulation, a relaxation parameter is introduced into the indicator function, and $\eta_{j}(\vect{x};\lambda _{j})$ is defined as
		\begin{equation}\label{elevenEq}
			\eta_{j}(\vect{x};\lambda _{j})=\mathbb{I}(G(\vect{x})\leqslant \lambda _{j})f_{\vect{X}}(\vect{x}),\quad j=1,2,...,T\,,
		\end{equation}
		with $\lambda _{1}>\lambda _{2}>\cdots >\lambda _{T}=0$ and the optimal importance densities for relaxed failure events can be expressed as
		\begin{equation}\label{twelveEq}
			h_{j}^{\ast }(\vect{x};\lambda _{j})=\frac{\eta_{j}(\vect{x};\lambda _{j})}{P_{j}}=\frac{\mathbb{I}(G(\vect{x})\leqslant \lambda _{j})f_{\vect{X}}(\vect{x})}{\int _{\vect{x}\in \rn}\mathbb{I}(G(\vect{x})\leqslant \lambda _{j})f_{\vect{X}}(\vect{x})\mathrm{d}\vect{x}},\quad j=1,2,...,T-1\,.
		\end{equation}
		
		Substituting Eq.\eqref{elevenEq} and Eq.\eqref{twelveEq} into Eq.\eqref{tenthEq}, the failure probability can be expressed by
		\begin{equation}\label{thirteenEq}
			\begin{split}
				P_{f}&=P_{1}\prod _{j=1}^{T-1}\int _{\vect{x}\in \rn}\frac{\mathbb{I}(G(\vect{x})\leqslant \lambda _{j+1})}{\mathbb{I}(G(\vect{x})\leqslant \lambda _{j})}h_{j}^{\ast}(\vect{x};\lambda _{j})\mathrm{d}\vect{x}\\&=p^{T-1}\int _{\vect{x}\in \rn}\frac{\mathbb{I}(G(\vect{x})\leqslant \lambda _{T})}{\mathbb{I}(G(\vect{x})\leqslant \lambda _{T-1})}h_{T-1}^{\ast}(\vect{x};\lambda _{T-1})\mathrm{d}\vect{x}\,.
			\end{split}
		\end{equation}
		
		The last line of Eq.\eqref{thirteenEq} is from the algorithmic implementation of subset simulation: the relaxation parameter is adaptively specified as the $p$ percentile value of $G(\vect{x})$, thus $p=\int _{\vect{x}\in \rn}\frac{\mathbb{I}(G(\vect{x})\leqslant \lambda _{j+1})}{\mathbb{I}(G(\vect{x})\leqslant \lambda _{j})}h_{j}^{\ast}(\vect{x};\lambda _{j})\mathrm{d}\vect{x}$ for $j=1,2,...,T-2$. Therefore, the well-known subset simulation \cite{au2001estimation} is consistent with the RIS framework. Notably, the intermediate probabilities (Eq.\eqref{ninethEq}) of subset simulation correspond to a discretized cumulative distribution function (CDF) $F_{G}(\lambda_{j})=\mathbb{P}(G(\vect{X})\leqslant \lambda_{j})$ at points $\lambda_{j}$, $j=1,2,...,T$, with the sequence of relaxation parameter values being the discretization points. Therefore, subset simulation is particularly attractive if, other than the failure probability, the distribution $F_{G}(g)$ is also of interest.
		
		\subsection{Sequential importance sampling}\label{SIS}
		\noindent The sequential importance sampling introduces a relaxation parameter into a smooth approximation of the indicator function, and $\eta_{j}(\vect{x};\lambda _{j})$ is defined as
		\begin{equation}\label{fifteenEq}
			\eta_{j}(\vect{x};\lambda _{j})=\Phi\left(\frac{-G(\vect{x})}{\lambda _{j}}\right)f_{\vect X}(\vect x),\quad j=1,2,...,T\,,
		\end{equation}
		where $\lambda _{1}>\lambda _{2}>\cdots>\lambda _{T}=0$ and $\Phi(\cdot )$ is the standard normal CDF. To simplify subsequent notations, we define $\mathbb{I}(G(\vect x)\leq0)\equiv\Phi\left(\frac{-G(\vect{x})}{\lambda _{T}}\right)$ for the limiting case $\lambda _{T}=0$; this is from the property $\mathbb{I}(G(\vect x)\leq0)=\lim_{\lambda\to0}\Phi\left(\frac{-G(\vect{x})}{\lambda}\right)$. Similar to subset simulation, for a large $\lambda_j$, the intermediate probability associated with $\eta_{j}(\vect{x};\lambda _{j})$ is large; for $\lambda_j=0$, the original failure probability is restored. Note that other than $\Phi(\cdot )$, there are other schemes to smooth the indicator function \cite{katafygiotis2007estimation}.
		
		The optimal importance densities for relaxed failure events can be expressed as
		\begin{equation}\label{sixteenEq}
			h_{j}^{\ast }(\vect{x};\lambda _{j})=\frac{\eta_{j}(\vect{x};\lambda _{j})}{P_{j}}=\frac{\Phi\left(\frac{-G(\vect{x})}{\lambda _{j}}\right)f_{\vect{X}}(\vect{x})}{\int _{\vect{x}\in \rn}\Phi\left(\frac{-G(\vect{x})}{\lambda _{j}}\right)f_{\vect{X}}(\vect{x})\mathrm{d}\vect{x}},\quad j=1,2,...,T-1\,.
		\end{equation}
		
		Substitution of Eq.\eqref{fifteenEq} and Eq.\eqref{sixteenEq} into Eq.\eqref{tenthEq} yields the failure probability estimation:
		\begin{equation}\label{seventeenEq}
			P_{f}=P_{1}\prod _{j=1}^{T-1}\int _{\vect{x}\in \rn}\frac{\Phi\left(\frac{-G(\vect{x})}{\lambda _{j+1}}\right)}{\Phi \left(\frac{-G(\vect{x})}{\lambda _{j}}\right)}h_{j}^{\ast}(\vect{x};\lambda _{j})\mathrm{d}\vect{x}\,.
		\end{equation}
		
		Again, the sequential importance sampling \cite{papaioannou2016sequential} is consistent with the RIS framework. The intermediate probabilities (Eq.\eqref{ninethEq}) of sequential importance sampling do not have direct practical applications. However, since the binary indicator function has been smoothed by a differentiable function, the sequential importance sampling is desirable for reliability sensitivity analysis \cite{papaioannou2018reliability}.
		
		\subsection{Annealed importance sampling}\label{AIS}
		\noindent The annealed importance sampling introduces a relaxation parameter into the distribution function $f_{\vect{X}}(\vect{x})$, and $\eta_{j}(\vect{x};\lambda _{j})$ is defined as
		\begin{equation}\label{eighteenEq}
			\eta_{j}(\vect{x};\lambda _{j})=\mathbb{I}(G(\vect{x})\leqslant 0)f_{\vect{X}}(\vect{x};\vect{I}\lambda_{j}^{2}),\quad j=1,2,...,T\,,
		\end{equation}
		where $\vect{I}$ is the identity matrix, $\lambda _{1}>\lambda _{2}>\cdots >\lambda _{T}=1$, and $f_{\vect{X}}(\vect{x};\vect{I}\lambda_{j}^{2})$ is the multivariate zero-mean normal PDF with covariance matrix $\vect{I}\lambda_{j}^{2}$. A large $\lambda_j$ corresponds to a large dispersion for $f_{\vect{X}}(\vect{x};\vect{I}\lambda_{j}^{2})$, resulting in a high probability for the relaxed failure event. Accordingly, the optimal importance densities for relaxed failure events can be expressed as
		\begin{equation}\label{nineteenEq}
			h_{j}^{\ast }(\vect{x};\lambda _{j})=\frac{\eta_{j}(\vect{x};\lambda _{j})}{P_{j}}=\frac{\mathbb{I}(G(\vect{x})\leqslant 0)f_{\vect{X}}(\vect{x};\vect{I}\lambda_{j}^{2})}{\int _{\vect{x}\in \rn}\mathbb{I}(G(\vect{x})\leqslant 0)f_{\vect{X}}(\vect{x};\vect{I}\lambda_{j}^{2})\mathrm{d}\vect{x}},\quad j=1,2,...,T-1\,.
		\end{equation}
		Substituting Eq.\eqref{eighteenEq} and Eq.\eqref{nineteenEq} into Eq.\eqref{tenthEq}, we obtain the failure probability estimation:
		\begin{equation}\label{twentyEq}
			P_{f}=P_{1}\prod _{j=1}^{T-1}\int _{\vect{x}\in \rn}\frac{f_{\vect{X}}(\vect{x};\vect{I}\lambda_{j+1}^{2})}{f_{\vect{X}}(\vect{x};\vect{I}\lambda_{j}^{2})}h_{j}^{\ast}(\vect{x};\lambda _{j})\mathrm{d}\vect{x}\,.
		\end{equation}
		
		For a standard multivariate Gaussian distribution, modifying the covariance matrix by $\vect{I}\lambda_{j}^{2}$ is equivalent to raising $f_{\vect{X}}(\vect{x})$ to some power. Therefore, the annealed importance sampling introduced here is a special case of \cite{neal2001annealed}. Notably, in fragility analysis, the dispersion of the sampling distribution is often proportional to the intensity measure of a hazard \cite{kiureghian2009nonlinear,yi2019gaussian}; therefore, the intermediate probabilities (Eq.\eqref{ninethEq}) of the annealed importance sampling correspond to a discretized fragility curve. The implementation details of annealed importance sampling for reliability problems are introduced in Appendix \ref{Fig:AIS}, as to our knowledge, the  annealed importance sampling has not been thoroughly investigated in the structural reliability community. 
		
		To summarize Section \ref{Framework}, RIS provides a unified framework to formulate subset simulation, sequential importance sampling, and annealed importance sampling. It is observed that the selection of particular relaxation parameters facilitates specific applications. 
		Using the RIS framework, we will develop new variants in the subsequent section. Finally, it is worth noting that subset simulation and sequential importance sampling are applicable to high-dimensional problems \cite{papaioannou2016sequential} because their relaxation parameters are applied to the indicator function. The annealed importance sampling may be restricted to low-dimensional problems because the relaxation parameter is introduced into the PDF, and the Gaussian PDF values decrease exponentially with dimensionality \cite{katafygiotis2008geometric}. Nevertheless, it will be shown in Section \ref{Sec:Applicationone} that the annealed importance sampling can exhibit remarkable efficiency for low-dimensional problems. 
		
		\section{New sampling strategies in the framework of relaxation-based importance sampling}\label{Newvariant}
		\noindent Existing variance reduction techniques, within the framework of RIS, introduce a single relaxation parameter either to the indicator function or to the probability density function. To showcase the potential of RIS in generating new IS schemes, hereafter, we will develop two new IS methods, each with two relaxation parameters. These new variants can significantly outperform existing variance reduction techniques in fragility problems. 
		
		\subsection{Importance sampling strategy I }\label{ISONE}
\begin{figure}[t]
			\centering
			\includegraphics[scale=0.5]{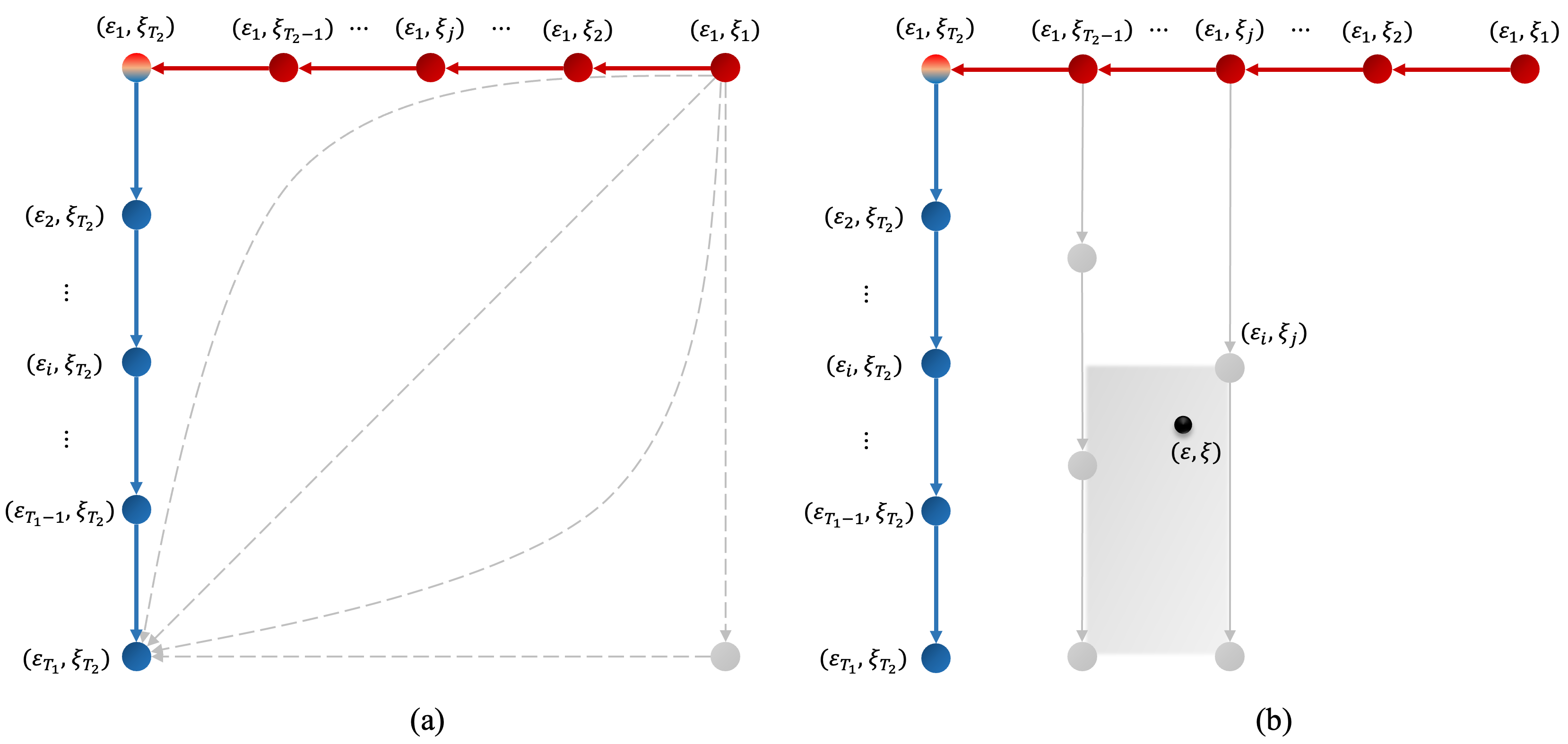}
			\caption{\textbf{(a) different routes to tune relaxation parameters  $\varepsilon_{i}$ and $\xi_{j}$, and (b) techniques to obtain a smooth fragility surface}. \textit{Panel (a): the solid red/blue arrows represent the adopted route to tune $(\varepsilon,\xi)$ values. The dashed arrows illustrate some other possible paths. Panel (b): at each red node, we have intermediate results regarding the failure probability, importance density, and samples (including their limit-state function values) from the importance density. Starting from these intermediate results, we perform a RIS in the direction of the gray arrows. Using the results collected at the gray nodes, i.e., grid points for fragility surface interpolation, in conjunction with Eq.\eqref{Newonefragilitysurface}, we can, without additional limit-state function evaluations,  estimate the failure probability for any point within $[\varepsilon _{T_1},\varepsilon _{1}]\times[\xi_{T_2},\xi_{1}]$, i.e., constructing a smooth fragility surface.}}
			\label{Fig:Illustration}
		\end{figure}

		\noindent We introduce two relaxation parameters, $\varepsilon _{T_1}$ and $\xi_{T_2}$, into the indicator and probability density functions, respectively, and $\eta_{T}(\vect{x};\vect{\lambda }_{T})$ can be written as
		\begin{equation}\label{Newonetarfaievent}
			\eta_{T}(\vect{x};\vect{\lambda }_{T})=\mathbb{I}(G(\vect{x})\leqslant\varepsilon _{T_1})f_{\vect{X}}(\vect{x};\vect{I}\xi_{T_2}^{2})
		\end{equation}
		where $\vect{\lambda }_{T}=[\varepsilon_{T_1},\xi_{T_2}]=[0,1]$ so that $\eta_{T}(\vect{x};\vect{\lambda }_{T})=\mathbb{I}(G(\vect{x})\leqslant0)f_{\vect{X}}(\vect{x})$. Tuning up $\varepsilon_{T_1}$ relaxes the event $\lbrace\vect x\in\rn:G(\vect x)\leq\varepsilon_{T_1}\rbrace$ and thus increases the failure probability, while tuning up $\xi_{T_2}$ increases the dispersion of $f_{\vect X}(\vect x;\vect{I}\xi_{T_2}^{2})$ and also increases the failure probability. Therefore, different ``routes" to tune the relaxation parameters are possible. We consider a simple two-step route (Figure \ref{Fig:Illustration} (a)) to tune the two relaxation parameters. Specifically, we start from $(\varepsilon_{1},\xi_{1})$ such that the probability associated with $\mathbb{I}(G(\vect{x})\leqslant\varepsilon _{1})f_{\vect{X}}(\vect{x};\vect{I}\xi_{1}^{2})$ is sufficiently large. Next, we fix $\varepsilon_{1}$ and gradually tune $\xi_{1}$ back to $\xi_{T_2}$. Finally, we fix $\xi_{T_2}$ and gradually tune $\varepsilon_{1}$ back to $\varepsilon_{T_1}$. {The general principle for tuning the relaxation parameter values is that the successive changes should balance the accuracy and efficiency, so that $(\varepsilon_{T_1},\xi_{T_2})$ is reached fast, and the importance sampling at each step is accurate. In algorithmic implementations, this general principle is materialized as setting a target probability for the sequence of relaxed failure events. Similar techniques are studied in \cite{del2006sequential,papaioannou2016sequential}.}	
		Provided with the route to tune relaxation parameters, $\eta_{j}(\vect{x};\vect{\lambda }_{j})$ can be formulated as
		
		\begin{equation}\label{Newonerelaxfaievent}
			\eta_{j}(\vect{x};\vect{\lambda }_{j})=\left\{\begin{aligned}
				&\mathbb{I}(G(\vect{x})\leqslant\varepsilon _{1})f_{\vect{X}}(\vect{x};\vect{I}\xi_{j}^{2}),&& j=1,2,\cdots ,T_2\\
				&\mathbb{I}(G(\vect{x})\leqslant\varepsilon _{j-T_2+1})f_{\vect{X}}(\vect{x};\vect{I}\xi_{T_2}^{2}), 
				&& j=T_2+1,T_2+2,\cdots ,T_2+T_1-1
			\end{aligned}\right.
		\end{equation}
		where $\varepsilon _{1}>\varepsilon _{2}>\cdots >\varepsilon _{T_1}=0$ and $\xi _{1}>\xi _{2}>\cdots >\xi _{T_2}=1$.

		The optimal importance densities for relaxed failure events can be expressed as
		\begin{equation}\label{Newoneoptimaldensity}
			h_{j}^{\ast}(\vect{x};\vect{\lambda} _{j})=\frac{\eta_{j}(\vect{x};\vect{\lambda} _{j})}{P_j}=\left\{\small\begin{aligned}
				&\frac{\mathbb{I}(G(\vect{x})\leqslant\varepsilon _{1})f_{\vect{X}}(\vect{x};\vect{I}\xi_{j}^{2})}{\int _{\vect{x}\in \rn}\mathbb{I}(G(\vect{x})\leqslant\varepsilon _{1})f_{\vect{X}}(\vect{x};\vect{I}\xi_{j}^{2})\mathrm{d}\vect{x}},&&j=1,2,...,T_2 \\ 
				&\frac{\mathbb{I}(G(\vect{x})\leqslant\varepsilon_{j-T_2+1})f_{\vect{X}}(\vect{x};\vect{I}\xi_{T_2}^{2})}{\int _{\vect{x}\in \rn}\mathbb{I}(G(\vect{x})\leqslant\varepsilon _{j-T_2+1})f_{\vect{X}}(\vect{x};\vect{I}\xi_{T_2}^{2})\mathrm{d}\vect{x}},&& j=T_2+1,T_2+2,...,T_2+T_1-2
			\end{aligned}\right.
		\end{equation}
		Substitution of Eq.\eqref{Newonerelaxfaievent} and Eq.\eqref{Newoneoptimaldensity} into Eq.\eqref{tenthEq} yields the failure probability estimation:
		\begin{equation}\label{Newonefailureprob}
			P_{f}=P_{1}\prod _{j=1}^{T_2-1}\int _{\vect{x}\in \rn}\frac{f_{\vect{X}}(\vect{x};\vect{I}\xi_{j+1}^{2})}{f_{\vect{X}}(\vect{x};\vect{I}\xi_{j}^{2})}h_{j}^{\ast}(\vect{x};\vect{\lambda} _{j})\mathrm{d}\vect{x}\prod _{j=T_2}^{T_2+T_1-2}\int _{\vect{x}\in \rn}\frac{\mathbb{I}(G(\vect{x})\leqslant\varepsilon_{j-T_2+2})}{\mathbb{I}(G(\vect{x})\leqslant\varepsilon_{j-T_2+1})}h_{j}^{\ast}(\vect{x};\vect{\lambda} _{j})\mathrm{d}\vect{x}
		\end{equation}
		
		This IS strategy, IS-I for simplicity, couples annealed importance sampling with subset simulation. In fragility analysis, given a performance threshold $\varepsilon_{i}$, the set of probabilities from varying $\xi_{j}$ corresponds to a fragility curve; given an intensity $\xi_{i}$, the set of probabilities from varying $\varepsilon_{j}$ corresponds to multi-state failure probabilities. The Cartesian product of the two sets of probabilities yields a \textit{fragility surface}, i.e., failure probability as a function of intensity measure and performance state, a critical component in performance-based engineering \cite{porter2007creating,gunay2013peer}. {IS-I is designed to efficiently estimate the fragility surface (Figure \ref{Fig:Illustration} (b)). Specifically, at each red node, we have intermediate results regarding the failure probability, importance density, and samples (including their limit-state function values) from the importance density. Starting from these intermediate results, we perform RIS in the direction of the gray arrows to estimate the probabilities for gray nodes, where limit-state function evaluations are inevitable. The results from this step define a mesh with grid point values for the fragility surface. Thereafter, we can infer (see the equation below) the failure probability for \textbf{any} point within the mesh without limit-state function evaluations, i.e., produce a smooth fragility surface.} 
  
  Provided with the probability $P_{i,j}$ associated with $\eta_{i,j}(\vect{x};\varepsilon_i, \xi_j )=\mathbb{I}(G(\vect{x})\leqslant\varepsilon _{i})f_{\vect{X}}(\vect{x};\vect{I}\xi_{j}^{2})$, the failure probability for a nearby point $\eta(\vect{x};\varepsilon, \xi )$ can be estimated as
		\begin{equation}\label{Newonefragilitysurface}
			\begin{aligned}
				P(\varepsilon, \xi)=&P_{i,j}\int _{\vect{x}\in \rn}\frac{\mathbb{I}(G(\vect{x})\leqslant\varepsilon)f_{\vect{X}}(\vect{x};\vect{I}\xi^{2})}{\mathbb{I}(G(\vect{x})\leqslant\varepsilon _{i})f_{\vect{X}}(\vect{x};\vect{I}\xi_{j}^{2})}h_{i,j}^{\ast}(\vect{x};\varepsilon_{i},\xi_{j})\mathrm{d}\vect{x}\\
				\approx&\frac{P_{i,j}}{N}\sum_{k=1}^N\frac{\mathbb{I}(G(\vect{x}^{(k)})\leqslant\varepsilon)f_{\vect{X}}(\vect{x}^{(k)};\vect{I}\xi^{2})}{\mathbb{I}(G(\vect{x}^{(k)})\leqslant\varepsilon _{i})f_{\vect{X}}(\vect{x}^{(k)};\vect{I}\xi_{j}^{2})}\,,
			\end{aligned}
		\end{equation}
		where $\varepsilon\in[\varepsilon _{i+1},\varepsilon _{i}]$,  $\xi\in[\xi_{j+1},\xi_{j}]$, and $\vect{x}^{(k)}$ are random samples from $h_{i,j}^{\ast}(\vect{x};\varepsilon_{i},\xi_{j})$. This probability estimation comes for free because the random samples $\vect{x}^{(k)}$ and their $G(\vect{x}^{(k)})$ values can be inherited from the step of estimating $P_{i,j}$. This indicates that for each $(\varepsilon_i, \xi_j)$ one can choose arbitrarily many $(\varepsilon,\xi)$ points within $[\varepsilon _{i+1},\varepsilon _{i}]\times[\xi_{j+1},\xi_{j}]$ to estimate their failure probabilities, resulting in a smooth fragility surface. 
		
		{Notice that an importance sampling strategy similar to IS-I has been recently developed in \cite{cheng2022estimation}, in which they seek to find an optimal route from $(\varepsilon_{1},\xi_{1})$ to $(\varepsilon_{T_1},\xi_{T_2})$ through simultaneous tuning of two relaxation parameters. The difference is that IS-I aims to produce a smooth fragility surface rather than (only) estimating the probability of a failure event. In this context, a relatively dense ``mesh” for the design space of the two relaxation parameters is needed. If the mesh is too coarse, the fragility surface will not be accurate/smooth. It follows that the two-step route (Figure \ref{Fig:Illustration} (a)) adopted in this study is, in fact, the longest route, i.e., the most ``inefficient” route (in the sense that if one is only interested in a point-wise estimation rather than a surface), so that the mesh determined by this route is relatively dense.}
		
		If an annealed importance sampling is iterated for multiple performance thresholds, or a subset simulation is iterated for multiple intensity measures, one can obtain a discretized fragility surface. In comparison, the strength of the proposed coupled strategy is performing importance sampling in \textbf{both} directions, leading to higher efficiency and a smooth fragility surface. The limitation of the proposed IS-I is the scalability to high-dimensional problems, since a relaxation parameter is introduced to the PDF, which decays exponentially with dimensionality. This limitation motivates the development of the following method,  leveraging a spherical formulation. Finally, the implementation details of IS-I for fragility surface estimation are summarized in Appendix \ref{Fig:ISone}. 
		
		\subsection{Importance sampling strategy II }\label{ISTWO}
		\noindent The failure probability defined in the $n$-dimensional standard normal space admits a spherical formulation:
		\begin{equation}\label{Sphericalrepresentation}
			P_f=\int _{r> 0}\int _{\vect{{u}}\in S^{n-1}}\mathbb{I}(G(\vect{{u}}\cdot r)\leqslant0)f_{U}(\vect{{u}})f_{\chi }(r)\mathrm{d}\vect{{u}}\mathrm{d}r\,,
		\end{equation}
		where $S^{n-1}$ denotes the ($n-1$)-dimensional unit hypersphere, $\vect{{u}}$ denotes a unit vector on $S^{n-1}$, $f_{U}(\vect{{u}})$ is the constant PDF of the uniform distribution on $S^{n-1}$, $r$ denotes a distance from the origin, $f_{\chi }(r)$ is the PDF of the $\chi$-distribution with $n$ degrees of freedom, and $\mathbb{I}(G(\vect{u}\cdot r)\leqslant0)$ is the binary indicator function that gives 1 if $G(\vect{u}\cdot r)\leqslant0$ and 0 otherwise.
		
		We introduce two relaxation parameters, $\varepsilon _{T_1}$ and $\xi_{T_2}$, both into the indicator function, and $\eta_{T}(\vect{u},r;\vect{\lambda }_{T})$ can be written as
		\begin{equation}\label{Newtwotarfaievent}
			\eta_{T}(\vect{u},r;\vect{\lambda }_{T})=\mathbb{I}(G(\vect{u}\cdot \xi_{T_2}r)\leqslant \varepsilon _{T_1})f_{U}(\vect{u})f_{\chi }(r)
		\end{equation}
		where $\vect{\lambda }_{T}=[\varepsilon_{T_1},\xi_{T_2}]=[0,1]$ so that $\eta_{T}(\vect{u},r;\vect{\lambda }_{T})=\mathbb{I}(G(\vect{{u}}\cdot r)\leqslant0)f_{U}(\vect{{u}})f_{\chi }(r)$. Similar to IS-I, tuning up $\varepsilon_{T_1}$ relaxes the failure event and thus increases the failure probability; while different from IS-I, tuning up $\xi_{T_2}$ projects $\vect u$ to a larger hypersphere, which is equivalent to increasing the sampling radius, and thus also increases the failure probability (similar ideas can be seen in \cite{wang2016cross,wang2018hyper}). We adopt the same route as in IS-I (see Figure \ref{Fig:Illustration} (a)) to tune the relaxation parameters. Accordingly, $\eta_{j}(\vect{u},r;\vect{\lambda }_{j})$ has the form
		\begin{equation}\label{Newtworelaxfaievent}
			\eta_{j}(\vect{u},r;\vect{\lambda }_{j})=\left\{\begin{aligned}
				&\mathbb{I}(G(\vect{u}\cdot \xi_{j}r)\leqslant\varepsilon _{1})f_{U}(\vect{u})f_{\chi }(r),&&j=1,2,...,T_2\\
				&\mathbb{I}(G(\vect{u}\cdot \xi_{T_2}r)\leqslant\varepsilon _{j-T_2+1})f_{U}(\vect{u})f_{\chi }(r),
				&&j=T_2+1,T_2+2,...,T_2+T_1-1
			\end{aligned}\right.	
		\end{equation}
		where $\varepsilon _{1}>\varepsilon _{2}>\cdots >\varepsilon _{T_1}=0$ and $\xi _{1}>\xi _{2}>\cdots >\xi _{T_2}=1$.
		
		The optimal importance densities for relaxed failure events can be expressed as
		\begin{equation}\label{Newtwooptimaldensity}
			h_{j}^{\ast}(\vect{u},r;\vect{\lambda} _{j})=\frac{\eta_{j}(\vect{u},r;\vect{\lambda} _{j})}{P_j}=\left\{\scriptsize\begin{aligned}
				&\frac{\mathbb{I}(G(\vect{u}\cdot \xi_{j}r)\leqslant\varepsilon _{1})f_{U}(\vect{u})f_{\chi }(r)}{\int _{r> 0}\int _{\vect{u}\in S^{n-1}}\mathbb{I}(G(\vect{u}\cdot \xi_{j}r)\leqslant\varepsilon _{1})f_{U}(\vect{u})f_{\chi }(r)\mathrm{d}\vect{u}\mathrm{d}r}\,,j=1,2,\cdots ,T_2 \\ 
				&\frac{\mathbb{I}(G(\vect{u}\cdot \xi_{T_2}r)\leqslant\varepsilon _{j-T_2+1})f_{U}(\vect{u})f_{\chi }(r)}{\int _{r> 0}\int _{\vect{u}\in S^{n-1}}\mathbb{I}(G(\vect{u}\cdot \xi_{T_2}r)\leqslant\varepsilon _{j-T_2+1})f_{U}(\vect{u})f_{\chi }(r)\mathrm{d}\vect{u}\mathrm{d}r}\,,j=T_2+1,...,T_2+T_1-2
			\end{aligned}\right.
		\end{equation}
		Substituting Eq.\eqref{Newtworelaxfaievent} and Eq.\eqref{Newtwooptimaldensity} into Eq.\eqref{tenthEq}, the failure probability can be obtained as
		\begin{equation}\label{Newtwofailureprob}
			\begin{split}
				P_{f}&=P_{1}\prod _{j=1}^{T_2-1}\int _{r> 0}\int _{\vect{u}\in S^{n-1}}\frac{\mathbb{I}(G(\vect{u}\cdot \xi_{j+1}r)\leqslant\varepsilon _{1})}{\mathbb{I}(G(\vect{u}\cdot \xi_{j}r)\leqslant\varepsilon _{1})}h_{j}^{\ast}(\vect{u},r;\vect{\lambda} _{j})\mathrm{d}\vect{u}\mathrm{d}r\\&\prod _{j=T_2}^{T_2+T_1-2}\int _{r> 0}\int _{\vect{u}\in S^{n-1}}\frac{\mathbb{I}(G(\vect{u}\cdot \xi_{T_2}r)\leqslant\varepsilon_{j-T_2+2})}{\mathbb{I}(G(\vect{u}\cdot \xi_{T_2}r)\leqslant\varepsilon_{j-T_2+1})}h_{j}^{\ast}(\vect{u},r;\vect{\lambda} _{j})\mathrm{d}\vect{u}\mathrm{d}r\,.
			\end{split}
		\end{equation}
		
		This IS strategy, IS-II for simplicity, is desirable for high-dimensional fragility problems. The relaxation parameter $\xi_j$ is a scaling factor for $\vect u$ and is proportional to the intensity measure. Therefore, similar to IS-I, IS-II can be adapted to efficiently estimate the fragility surface. Compared with IS-I, the tiny but crucial difference is that IS-II introduces both relaxation parameters into the indicator function, leveraging a spherical formulation. Consequently, IS-II applies to high-dimensional problems. 
		
		The benefit of introducing $\xi_j$ into the indicator function--limit-state function, more specifically--comes with a price: determining a proper value for $\xi_j$ now involves limit-state function evaluations. Note that, in IS-I, testing with trial values of $\xi_j$ does not involve limit-state function evaluations, because $\xi_j$ is a parameter of the PDF $f_{\vect X}(\vect x;\vect I \xi_j^2)$. Given a current $\xi_{j}$, determining the next $\xi_{j+1}$, in principle, requires solving the optimization problem: 
		\begin{equation}\label{solverelaxpara}
			\xi_{j+1}=\mathop{\arg\min}\limits_{\xi\in [1,\xi_{j})}\left \| \int _{r> 0}\int _{\vect{u}\in S^{n-1}}\frac{\mathbb{I}(G(\vect{u}\cdot \xi r)\leqslant\varepsilon _{1})}{\mathbb{I}(G(\vect{u}\cdot \xi_{j}r)\leqslant\varepsilon _{1})}h_{j}^{\ast}(\vect{u},r;\vect{\lambda} _{j})\mathrm{d}\vect{u}\mathrm{d}r-\rho\right \|\,,
		\end{equation}
		where $\rho$ is a parameter controlling the target value of intermediate probabilities. It is seen from Eq.\eqref{solverelaxpara} that testing with trial values for $\xi$ involves reevaluating $G(\vect u\cdot\xi r)$. To avoid this computation, we have developed an extrapolation technique to predict the next $\xi_{j+1}$ using results from previous steps. Furthermore, using the extrapolation technique in conjunction with an approach similar to Eq.\eqref{Newonefragilitysurface}, we have developed a scheme to interpolate a smooth fragility surface. For simplicity and readability, we discuss technical details regarding the relaxation parameter extrapolation and fragility surface interpolation in \ref{App: Extrapolation}. The main message from \ref{App: Extrapolation} is that in contrast with Eq.\eqref{Newonefragilitysurface} of IS-I, which is theoretically rigorous, the interpolation process of IS-II involves approximations to avoid additional limit-state function calls. However, similar to IS-I, the grid points for interpolation are obtained from importance sampling and thus have guaranteed accuracy. The implementation details of IS-II for estimating the fragility surface is summarized in Appendix \ref{Fig:IStwo}. 
		
		To briefly summarize Section \ref{Newvariant}, two variants of RIS, IS-I and IS-II, are developed to produce the fragility surface for performance-based engineering. The IS-I is designed for low-dimensional problems, while IS-II is tailored for high-dimensional problems. Their performance will be demonstrated in Section \ref{Application}. Finally, although theoretically trivial, it is worth mentioning that if a single fragility curve is of interest rather than the entire fragility surface, one can construct special cases of IS-I and IS-II by simply fixing the relaxation parameter $\varepsilon$. In this context, IS-I reduces to the annealed importance sampling, and IS-II reduces to an adaptive spherical importance sampling method based on radius scaling, which is still a novel importance sampling formulation. 
		
		{It is worth noting that the idea of scaling the input variance traces back to the work in \cite{bucher2009asymptotic}. Recently, this idea was used in developing a variant of subset simulation \cite{rashki2021sesc} and sequential directional importance sampling \cite{cheng2023rare}. In \cite{rashki2021sesc}, a Cartesian formulation is adopted and a relaxation parameter is introduced into the limit-state function, while \cite{cheng2023rare} adopts a spherical formulation and applies a relaxation parameter to the PDF of the $\chi$-distribution. These two  sampling methods are consistent with RIS using a single relaxation parameter, aiming for a point-wise failure probability estimation.}
				
		\section{Numerical examples}\label{Application}
		\noindent As the literature lacks studies of annealed importance sampling for reliability problems, this section first investigates a 2-dimensional reliability problem with a parabolic limit-state function to demonstrate the performance of annealed importance sampling. Subsequently, a 2-dimensional and 1000-dimensional seismic fragility problem are studied to illustrate the performance of the proposed IS-I and IS-II, respectively. For all RIS methods considered, the Hamiltonian Monte Carlo \cite{neal2011mcmc,wang2019hamiltonian} is adopted for sampling from the intermediate importance densities.
		
		\subsection{2-dimensional parabolic limit-state function}\label{Sec:Applicationone}
		\noindent Consider a reliability problem with a parabolic limit-state function:
		\begin{equation}\label{gfunction}
			G(\vect{X})=d-X_2-0.5(X_1-0.1)^2\,,
		\end{equation}
		where $X_1$ and $X_2$ are independent standard normal random variables,  and $d$ controls the failure probability. Three cases corresponding to $d=5$, $d=7$, and $d=9$ are studied, and the annealed importance sampling is compared with the Hamiltonian Monte Carlo-based Subset Simulation \cite{wang2019hamiltonian}. The direct Monte Carlo simulation with $10^8$ samples is used to obtain the reference solutions. The results are reported in Table \ref{tab:1}. It is observed that, for different failure probabilities, both the annealed importance sampling and subset simulation match the reference solution, but the annealed importance sampling is noticeably more efficient, especially for small failure probabilities.
		
\begin{figure}[H]
			\centering
			\includegraphics[scale=0.5]{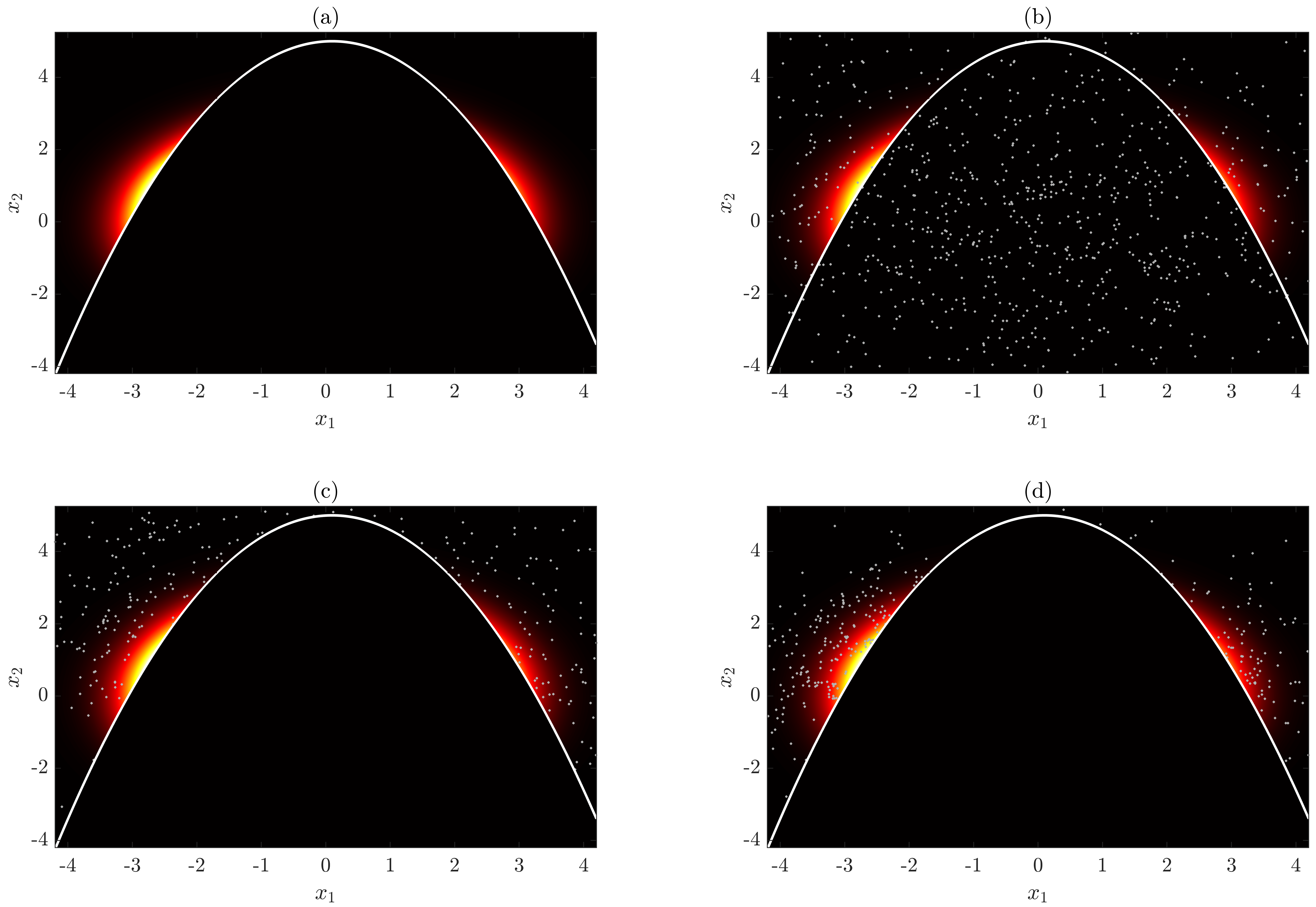}
			\caption{\textbf{Optimal importance density and samples of intermediate importance  densities.} \textit{(a) optimal importance density; (b) initial samples from $f_{\vect X}(\vect x,\vect I\lambda_1^2)$; (c) samples from $h_{1}^{\ast}(\vect{x};\lambda _{1})$; (d) samples from $h_{2}^{\ast}(\vect{x};\lambda _{2})$. The annealed importance sampling takes three iterations to estimate the failure probability. The samples are highly dispersed initially, and they become more concentrated around the high-density region of the failure domain as the algorithm proceeds}.}
			\label{Fig:Applicationone}
		\end{figure}

		For $d=5$, the optimal importance density of $G(\vect{x})$ is illustrated in Figure \ref{Fig:Applicationone} (a), and the iterative process of annealed importance sampling is shown in Figure \ref{Fig:Applicationone} (b), (c), and (d). It is seen from Figure \ref{Fig:Applicationone} (b) that at the initial step of annealed importance sampling, samples of $f_{\vect X}(\vect x,\vect I\lambda_1^2)$ already have a significant amount in the failure domain, due to the increase of dispersion. Furthermore, as
		the relaxed failure events approach the target failure event, the samples from the intermediate importance densities resemble the shape of the optimal importance density.
		
		\begin{table}[H]
			\centering
			\caption{Failure probability estimations from annealed importance sampling and subset simulation.}
			\begin{tabular}{c c c c c}
				\toprule
				Case  & Method & $\bar{N}$ & $\bar{P_f}$ & $\delta_{P_f}$\\ \midrule

				$d=5$ & Annealed importance sampling & 2800 & $3.00\times 10^{-3}$ & 0.1438 \\
				$(P_f=3.02\times 10^{-3})$ & Subset simulation & 2800 & $3.05\times 10^{-3}$ & 0.2301 \\
				$d=7$ & Annealed importance sampling & 2800 & $3.48\times 10^{-4}$ & 0.1715 \\
				$(P_f=3.46\times 10^{-4})$ & Subset simulation & 3700 & $3.45\times 10^{-4}$ & 0.2848 \\
				$d=9$ & Annealed importance sampling & 2800 & $4.20\times 10^{-5}$ & 0.1740 \\
				$(P_f=4.20\times 10^{-5})$ & Subset simulation & 4600 & $4.21\times 10^{-5}$ & 0.3488 \\
				\bottomrule
			\end{tabular}
			\label{tab:1}
		\end{table}
		
		\subsection{2-dimensional seismic fragility problem}\label{Sec:Applicationtwo}
		\noindent Consider a nonlinear hysteretic oscillator under seismic excitation with the equation of motion expressed as
		\begin{equation}\label{hystereticoscillator}
			m\ddot{u}(t)+c\dot{u}(t)+\alpha ku(t)+(1-\alpha)kz(t)=-m\ddot{u}_g(t)\,,
		\end{equation}
		where $m=3\times10^5\mathrm{kg}$, $c=5\times10^7\mathrm{N\cdot s/m}$, and $k=3\times10^7\mathrm{N/m}$ are the mass, damping, and initial stiffness of the oscillator, respectively, $\alpha=0.1$ is the stiffness reduction ratio, $u(t)$, $\dot{u}(t)$, and $\ddot{u}(t)$ are the displacement, velocity, and acceleration of the oscillator, respectively,
		$\ddot{u}_g(t)$ is the seismic acceleration excitation, and $z(t)$ is the hysteretic displacement governed by the nonlinear differential equation of Bouc-Wen model, expressed by \cite{wen1980equivalent}
		\begin{equation}\label{BoucWen}
			\dot{z}(t)=\phi \dot{u}(t)-\varphi \left |\dot{u}(t)  \right |z(t)\left |z(t)  \right |^{\gamma -1}-\psi  \dot{u}(t)\left |z(t)  \right |^{\gamma }\,,
		\end{equation}
		where $\phi=1$, $\varphi=\psi =400\mathrm{m}^{-1}$, and $\gamma=1$ are the shape parameters of the hysteresis loop, and the corresponding yield displacement is $u_y=(\varphi +\psi )^{-1/\gamma} =1.25\mathrm{mm}$.
		
		The seismic excitation is assumed to be a random combination of two ground motion  records, i.e.,
		\begin{equation}\label{Seismicexcitation}
			\ddot{u}_g(\vect{X},t)=X_1\ddot{u}_{g_1}(t)+X_2\ddot{u}_{g_2}(t)
		\end{equation}
		where $X_1$ and $X_2$ are two independent standard normal random variables, and $\ddot{u}_{g_1}(t)$ and $\ddot{u}_{g_2}(t)$ are the scaled N-S and E-W El-Centro records such that their peak ground accelerations (PGA) are 0.05g with g being the acceleration of gravity. The duration of the ground motion is $30$ seconds. 
		
		To define a seismic fragility problem, structural failure is defined as the absolute peak response exceeding a response threshold. The intensity measure is taken as the PGA, and its range is set as $[a_{l},a_{u}]=[0.05{\rm g},0.4{\rm g}]$. The range of the response threshold is set to $[b_l,b_u]=[u_y,2u_y]$. To initiate the IS-I algorithm, we set the failure event associated with $(a_{l},b_u)$, the lowest PGA and the highest response threshold, as the target failure event because it has the lowest failure probability. The failure event associated with $(a_{u},b_l)$, the highest PGA and the lowest response threshold, is the initial failure event with the highest failure probability. Consequently, the IS-I algorithm starts from $(a_{u},b_l)$ and ends at $[a_{l},b_u]$, estimating the intermediate failure probabilities within  $[a_{l},a_{u}]\times[b_{l},b_{u}]$ along the way, i.e., producing a fragility surface.

\begin{figure}[H]
			\centering
			\includegraphics[scale=0.4]{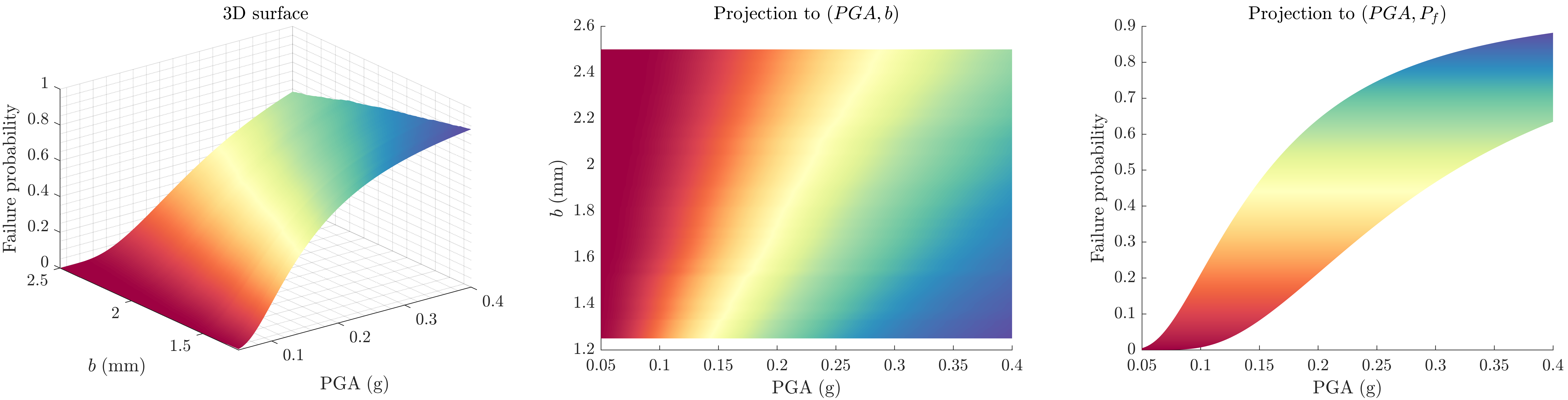}
			\caption{\textbf{Seismic fragility surface obtained from IS-I.} \textit{The fragility surface is produced by a single run of the proposed IS-I algorithm, using only 3700 limit-state function evaluations.}}
			\label{Fig:Applicationtwofragilitysurface}
		\end{figure}

\begin{figure}[H]
			\centering
			\includegraphics[scale=0.5]{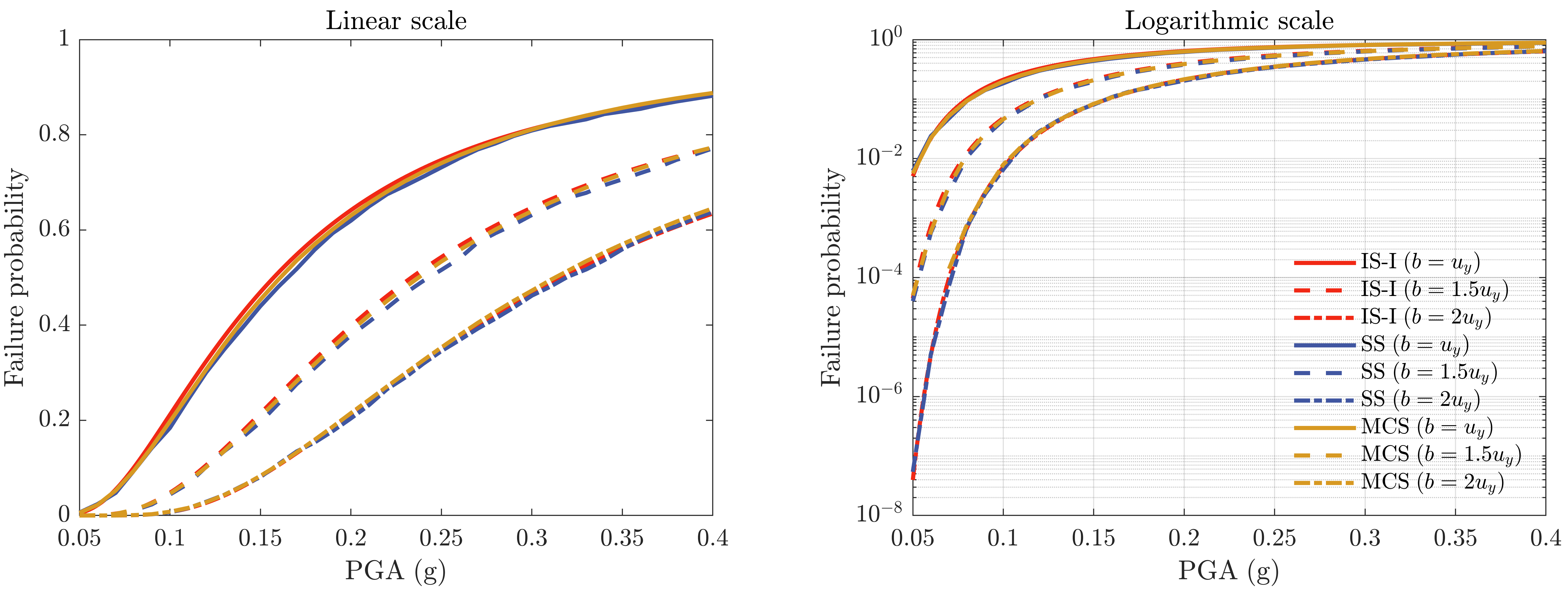}
			\caption{{\textbf{Seismic fragility curves obtained from IS-I.} \textit{(a) Linear scale; (b) Logarithmic scale. The fragility curves are obtained by simply reading the fragility surface with fixed performance thresholds. The subset simulation solutions iterated over various intensity measures using $6.12\times10^4$ samples,  and the direct Monte Carlo solutions using $3.6\times10^6$ samples are shown for comparisons. It is seen that the results from IS-I are in close agreement with the reference solutions}.}}
			\label{Fig:Applicationtwofragilitycurve}
		\end{figure}

		Simulating the IS-I algorithm once, a smooth seismic fragility surface is obtained, as shown in Figure \ref{Fig:Applicationtwofragilitysurface}. The fragility curves corresponding to thresholds $b=u_y$, $b=1.5u_y$, and $b=2u_y$, obtained by simply reading the fragility surface with fixed $b$, are presented in Figure \ref{Fig:Applicationtwofragilitycurve}. {Finally, it is remarkable to find out that IS-I only requires 3700 limit-state function calls to produce such a smooth fragility surface, {while the subset simulation iterated over various intensity measures requires $6.12\times10^4$ limit-state function calls to reach a similar accuracy, demonstrating the efficiency of the proposed IS-I for fragility surface estimation, i.e., the necessity of introducing two relaxation parameters.}}
		
		\subsection{1000-dimensional seismic fragility problem}\label{Sec:Applicationthree}
		\noindent We reconsider the previous example, but now the excitation is a stationary Gaussian white noise with a duration of $30$ seconds. Using the spectral representation method \cite{shinozuka1991simulation}, the white noise is described by a finite set of Gaussian variables:  
		\begin{equation}\label{Seismicexcitationrandomprocess}
			\ddot{u}_g(\vect{X},t)=\sum_{i=1}^{n/2}\sqrt{2S_{0}\Delta \omega }\left ( X_i\textrm{cos}(\omega_it)+ \bar{X}_i\mathrm{sin}(\omega_it)\right )
		\end{equation}
		where $X_i$ and $\bar{X}_i$, $i=1,2,...,n/2$, are mutually independent standard normal random variables, $n=1000$, $\Delta \omega=2\omega_{max}/n$ is the frequency increment with $\omega_{max}=25\pi$ being the upper cutoff
		angular frequency, $\omega_i=(i-0.5)\Delta\omega$ is the discretized frequency points, and $S_0=1.3\times10^{-4} \mathrm{m^2/s^3}$ is the intensity
		of the white noise such that the mean PGA is 0.05g.

\begin{figure}[H]
			\centering
			\includegraphics[scale=0.4]{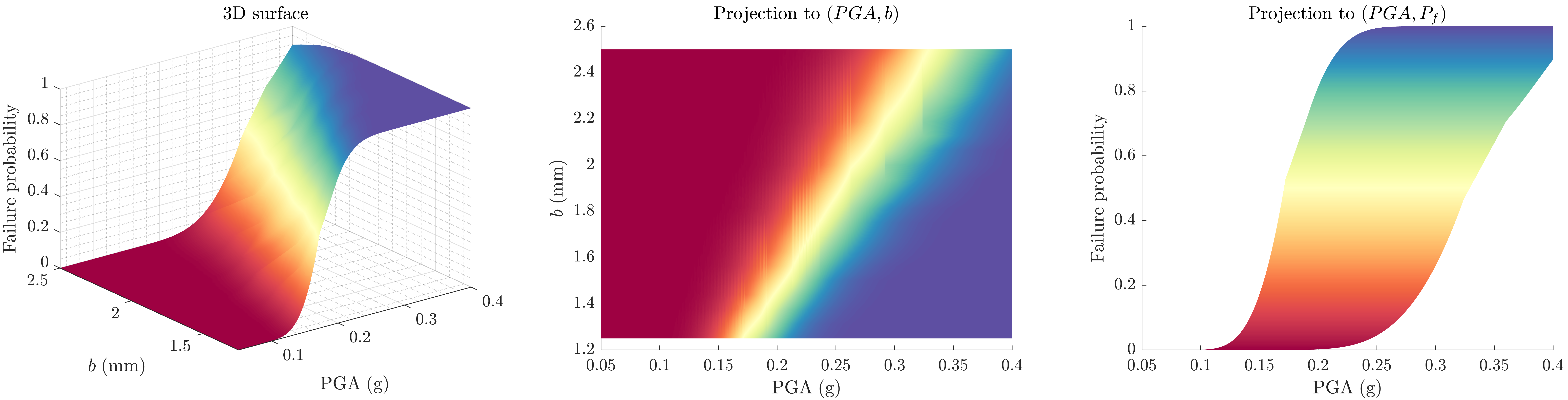}
			\caption{\textbf{Seismic fragility surface obtained from IS-II.} \textit{The fragility surface is produced by a single run of the proposed IS-II algorithm, using $3.2\times10^4$ limit-state function evaluations.}}
			\label{Fig:Applicationthreefragilitysurface}
		\end{figure}

\begin{figure}[H]
			\centering
			\includegraphics[scale=0.5]{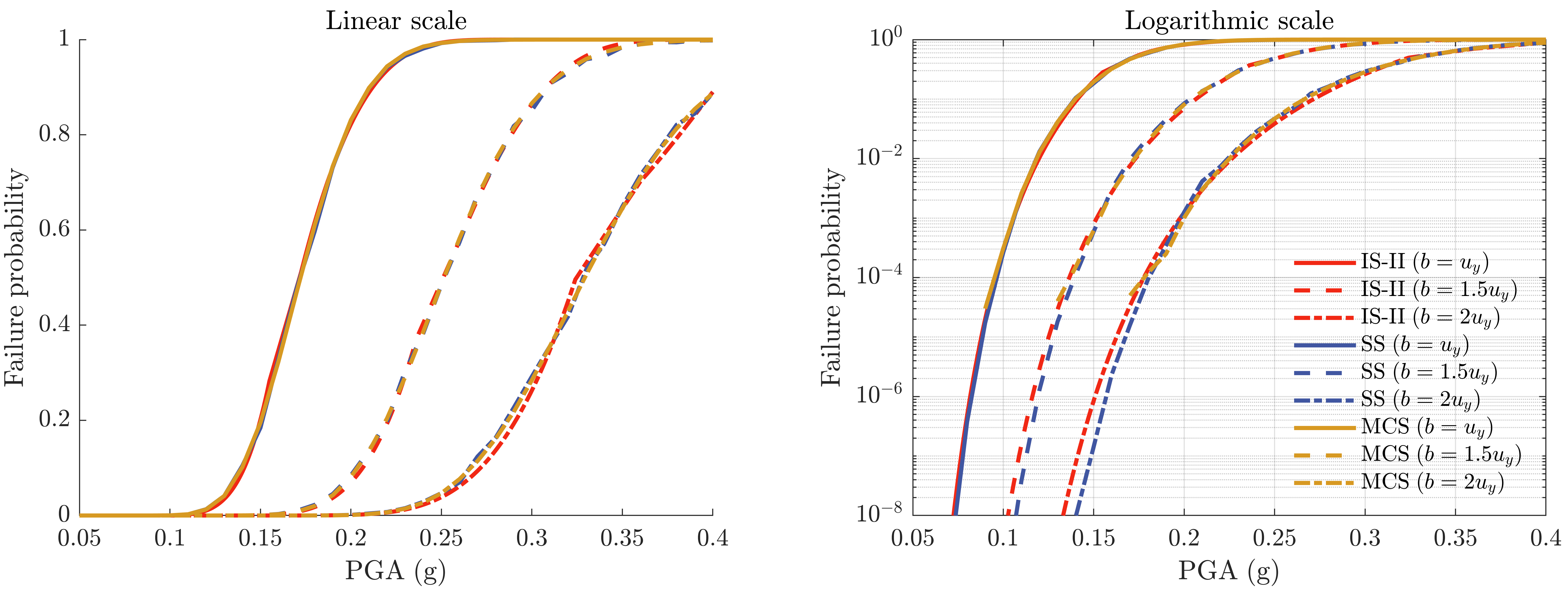}
			\caption{{\textbf{Seismic fragility curves obtained from IS-II.} \textit{(a) Linear scale; (b) Logarithmic scale. The fragility curves are obtained by simply reading the fragility surface with fixed performance thresholds. The subset simulation solutions iterated over various intensity measures using $2.88\times10^5$ samples, and the direct Monte Carlo solutions using $3.6\times10^6$ samples are shown for comparisons. It is seen that the results from IS-II are in close agreement with the reference solutions}.}}
			\label{Fig:Applicationthreefragilitycurve}
		\end{figure}

		The fragility problem is defined identically as the previous example. The seismic fragility surface obtained by IS-II is shown in Figure \ref{Fig:Applicationthreefragilitysurface}, and the seismic fragility curves corresponding to thresholds $b=u_y$, $b=1.5u_y$ and $b=2u_y$ are illustrated in Figure \ref{Fig:Applicationthreefragilitycurve}. {Again, the results from IS-II are in close agreement with the direct Monte Carlo simulation and subset simulation}. The IS-II algorithm takes $3.2\times10^4$ limit-state function evaluations to produce such a smooth fragility surface. This computational cost is one order larger than that of the IS-I for the previous low-dimensional fragility problem, because: i) IS-II introduces a relaxation parameter into the limit-state function and thus requires more limit-state function calls to estimate the intermediate failure probabilities (can be seen by comparing Eq.\eqref{Newtwofailureprob} with Eq.\eqref{Newonefailureprob}), and  ii) the current problem has smaller failure probabilities than the previous example. {The subset simulation iterated over various intensity measures requires $2.88\times10^5$ limit-state function calls to reach a similar accuracy.} Therefore, IS-II is still noticeably more efficient than other alternative algorithms. 
		
		{
			\section{Additional remarks and future directions}
			\subsection{Novelty of the relaxation-based importance sampling framework}
            \noindent RIS can be obtained by extracting a universal form from the sequential sampling methods, including sequential importance sampling, subset simulation, and annealed importance sampling. The sequential importance sampling studied in \cite{papaioannou2016sequential} offers a general form similar to RIS, although \cite{papaioannou2016sequential}  only investigates one possibility (recall Section \ref{SIS}) to introduce the relaxation parameter. Building on existing progress, the main message of this work is “a relaxation-based importance sampling strategy can be tailored to specific applications through a proper designing of relaxation parameters.” To our knowledge, this perspective has never been investigated. Moreover, the generalization of RIS offers a clear path to developing alternative importance sampling strategies tailored to different applications. 
			\subsection{How/where to introduce the relaxation parameters?}
			\noindent The specific strategy on how/where to introduce the relaxation parameters should be application-driven, i.e., one should specify the problem before designing the $\eta_{T}(\vect{x};\vect{\lambda } _{T})$ function and the corresponding relaxation parameters. For instance, if one is only interested in a CDF or fragility curve, a single relaxation parameter, e.g., using the subset simulation for the CDF and annealed importance sampling (or a special case of IS-II, i.e., fixing the performance threshold for high-dimensional stochastic excitation models) for the fragility curve, will suffice, because there is only one free variable in the CDF or fragility curve. The proposed IS-I and IS-II with two relaxation parameters are designed for estimating the fragility surface, which involves two free variables. In this context, introducing two relaxation parameters is ideal: using one parameter will be inefficient, while using three or more parameters will be redundant. 
			\subsection{Future directions}
			\begin{itemize}
				\item \textbf{Beyond two relaxation parameters}: A three or more-parameter family of RIS can be promising for multi-dimensional fragility analysis, e.g., multiple intensity measures are used to define the fragility function. 
				\item \textbf{Integration with sequential surrogate modeling}: In the context of surrogate modeling, the original, high-fidelity computational model can be progressively relaxed into a sequence of low-fidelity models, e.g., via adjusting the spatial and/or temporal meshes of the model. It will be promising to integrate RIS with the sequence of surrogate models and leverage the intermediate results.
				\item \textbf{Application in reliability-based design optimization}: In reliability-based design optimization, the design parameters can be treated analogously to the relaxation parameters. It is attractive to develop RIS techniques making full use of the intermediate probability estimations with respect to varying design parameters.
			\end{itemize}
			
		}
		\section{Conclusions}\label{conclude}
		\noindent This work proposes an importance sampling formulation based on adaptively relaxing parameters from the indicator function and/or the probability density function. By properly selecting the relaxation parameters, the relaxation-based importance sampling yields forms of subset simulation, sequential importance sampling, and annealed importance sampling. Moreover, the formulation lays the foundation for developing new importance sampling strategies, tailoring to specific applications. Two new importance sampling strategies are developed within the framework of relaxation-based importance sampling. The first strategy couples the annealed importance sampling with subset simulation, by introducing one relaxation parameter to the probability density function and another to the indicator function. The second strategy introduces two relaxation parameters into the indicator function, leveraging a spherical formulation for reliability problems in a Gaussian space. The first strategy is restricted to low-dimensional problems, while the second applies to high-dimensional scenarios. Both are desirable for fragility analysis in performance-based engineering, because they can be adapted to produce the entire fragility surface in a single run of the sampling algorithm. Three numerical examples are studied. The first example considers a 2-dimensional analytic nonlinear limit-state function to investigate the performance of annealed importance sampling in reliability analysis. The second and third examples are a 2-dimensional and a 1000-dimensional seismic fragility problem, designed to study the performance of the proposed new importance sampling strategies. It is found that the proposed methods are highly promising for fragility surface estimations, with computational efficiency outperforming existing sampling approaches. 
		
		
		\bibliography{EpidemicModel}
		
		\appendix
		
		\section{Technical details of IS-II for fragility surface estimation}\label{App: Extrapolation}
		\subsection*{Extrapolation to predict the next relaxation parameter}
		\noindent First, we replace the PDF $f_{\chi }(r)$ in Eq.\eqref{solverelaxpara} (inside the density $h_{j}^{\ast}(\vect{u},r;\vect{\lambda} _{j})$) by a Dirac delta function $\delta (r-R)$ with $R=\sqrt{n}$. This approximation is reasonable for high-dimensional problems, because the PDF $f_{\chi }(r)$ has high probability density in a small vicinity of $r=\sqrt{n}$ and the inner integral of Eq.\eqref{solverelaxpara} typically does not vary dramatically in that small vicinity (see \cite{wang2016cross} for detailed discussions). Accordingly, Eq.\eqref{solverelaxpara} can be rewritten as
		\begin{equation}\label{solverelaxparaappro}
			\xi_{j+1}=\mathop{\arg\min}\limits_{\xi\in [1,\xi_{j})}\left \| \frac{\theta (\xi R;\varepsilon _{1})}{\theta (\xi_{j}R;\varepsilon _{1})}-\rho\right \|\,,
		\end{equation}
		where $\theta (\xi R;\varepsilon _{1})=\int _{\vect{u}\in S^{n-1}}\mathbb{I}(G(\vect{u}\cdot \xi R)\leqslant\varepsilon _{1})f_{U}(\vect{u})\mathrm{d}\vect{u}$ is the \textit{failure ratio}, which has a clear geometric interpretation as the ratio between the hyperspherical surface area belonging to the failure domain and the total surface area of the hypersphere, given a radius $r=\xi R$. 
		
		Next, we adapt the extrapolation formula developed in \cite{wang2018hyper} to predict the failure ratio $\theta (\xi R;\varepsilon _{1})$ of Eq.\eqref{solverelaxparaappro}, without evaluating the limit-state function. The parameterized extrapolation model has the following form: 
		\begin{equation}\label{failureratiomodel1}
			\theta (r;\varepsilon _{1})\cong \hat{\theta} (r;\varepsilon _{1},\vect{v})=\frac{1}{2}\sum_{k=1}^{K}B\left ( 1-\left(\frac{b_k}{r}\right)^{2};\frac{n-1}{2} ,\frac{1}{2}\right ),\quad \theta (r;\varepsilon _{1})\in [0,0.5]\,,
		\end{equation}
		where $B(\cdot )$ is the incomplete beta function and $\vect{v}$ denotes a vector containing parameters $b_k$ and $K$, which can be determined through a nonlinear programming analysis based on data points of previous failure ratios. The failure ratio data points correspond to the intermediate failure probabilities of IS-II, applying the Dirac delta approximation for $f_{\chi }(r)$. The extrapolation model in Eq.\eqref{failureratiomodel1} is valid only when the failure ratio is within $[0,0.5]$. For the range $[0.5,1]$, we modify Eq.\eqref{failureratiomodel1} into
		\begin{equation}\label{failureratiomodel2}
			\theta (r;\varepsilon _{1})\cong \hat{\theta} (r;\varepsilon _{1},\vect{v})=1-\frac{1}{2}\sum_{k=1}^{K}B\left ( 1-\left(\frac{b_k}{a-r}\right)^{2};\frac{n-1}{2} ,\frac{1}{2}\right ),\quad \theta (r;\varepsilon _{1})\in [0.5,1]\,,
		\end{equation}
		where the vector $\vect{v}$ contains parameters $b_k$, $K$, and $a$, which can also be determined through a nonlinear programming analysis based on data points of previous failure ratios. Using the extrapolation models in Eq.\eqref{failureratiomodel1} and Eq.\eqref{failureratiomodel2}, we can solve the optimization Eq.\eqref{solverelaxparaappro} and  determine the relaxation parameter $\xi_{j+1}$, without evaluating the limit-state function.
		
		\subsection*{Fragility surface interpolation}
		\noindent Similar to IS-I, to obtain a smooth fragility surface, the counterpart of Eq.\eqref{Newonefragilitysurface} in the spherical formulation can be expressed as
		\begin{equation}\label{Newtwofragilitysurface}
			P(\varepsilon, \xi)=P_{i,j}\int _{r> 0}\int _{\vect{u}\in S^{n-1}}\frac{\mathbb{I}(G(\vect{u}\cdot \xi r)\leqslant\varepsilon)}{\mathbb{I}(G(\vect{u}\cdot \xi_j r)\leqslant\varepsilon_i)}h_{i,j}^{\ast}(\vect{u},r;\varepsilon_{i},\xi_{j})\mathrm{d}\vect{u}\mathrm{d}r\,.
		\end{equation}
		In contrast with Eq.\eqref{Newonefragilitysurface}, since the relaxation parameter $\xi$ is introduced into the limit-state function, the estimation of $P(\varepsilon, \xi)$ will not be free. To solve this issue, applying the Dirac delta approximation for $f_{\chi }(r)$, Eq.\eqref{Newtwofragilitysurface} is rewritten into
		\begin{equation}\label{Newtwofragilitysurfaceappro}
			P(\varepsilon, \xi)=P_{i,j}\frac{\theta (\xi R;\varepsilon)}{\theta (\xi_{j}R;\varepsilon _{i})}\,.
		\end{equation}
		Using the extrapolation models in Eq.\eqref{failureratiomodel1} and Eq.\eqref{failureratiomodel2}, we can predict $P(\varepsilon, \xi)$ without limit-state function evaluations. In contrast with IS-I, in IS-II, the scheme to interpolate a smooth fragility surface involves approximations. However, it is important to notice that the failure probabilities at grid points $(\varepsilon_i,\xi_j)$ are estimated by importance sampling and thus have theoretical guarantee of correctness. 
		
		\section{Implementation details}\label{Append:SolutionProcedureAIS}
		
		\begin{figure}[H]
			\centering
			\includegraphics[scale=0.8]{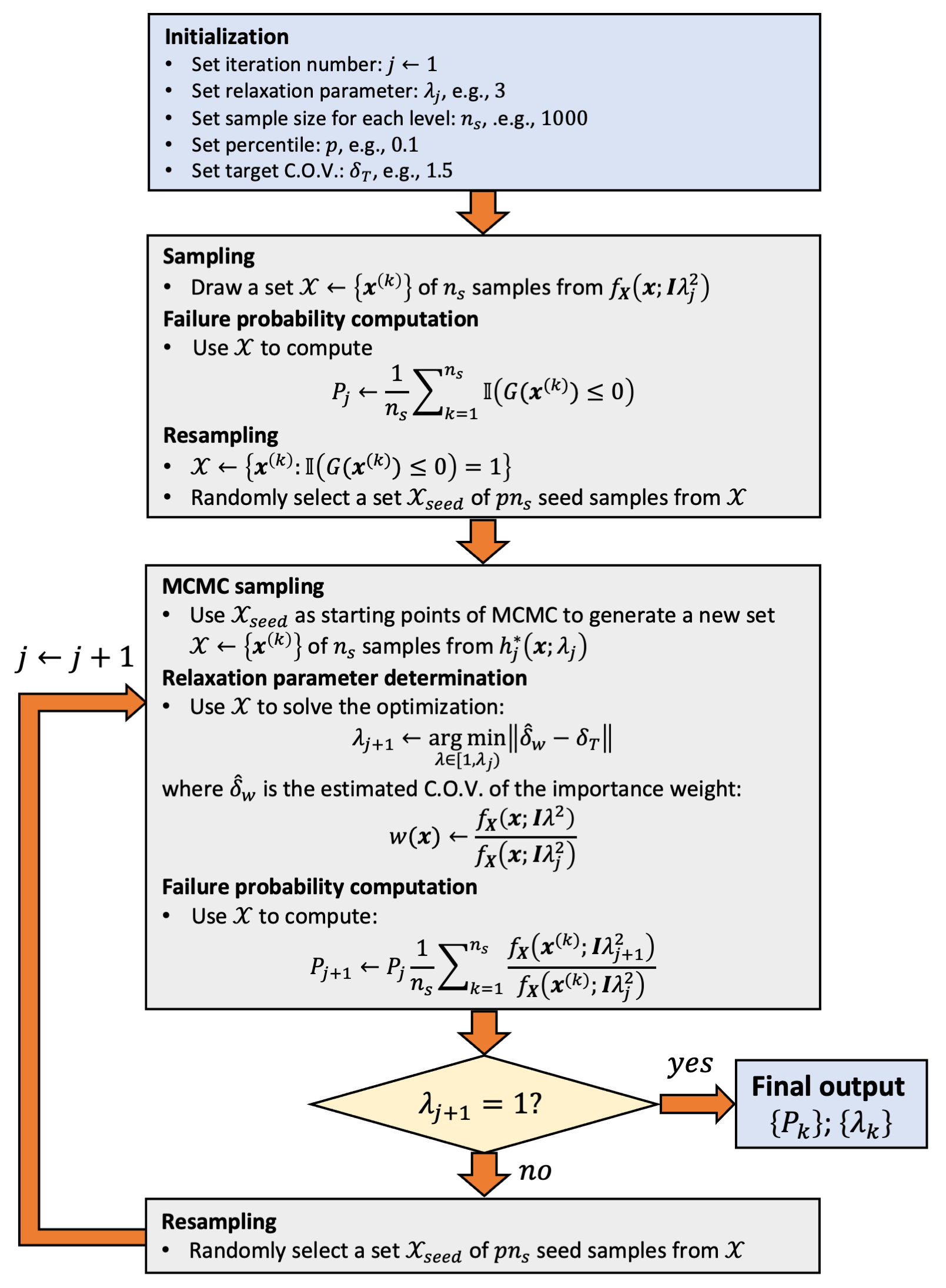}
			\caption{\textbf{Implementation details of annealed importance sampling for structural reliability analysis}.}
			\label{Fig:AIS}
		\end{figure}
		
		\begin{figure}[H]
			\centering
			\includegraphics[scale=0.7]{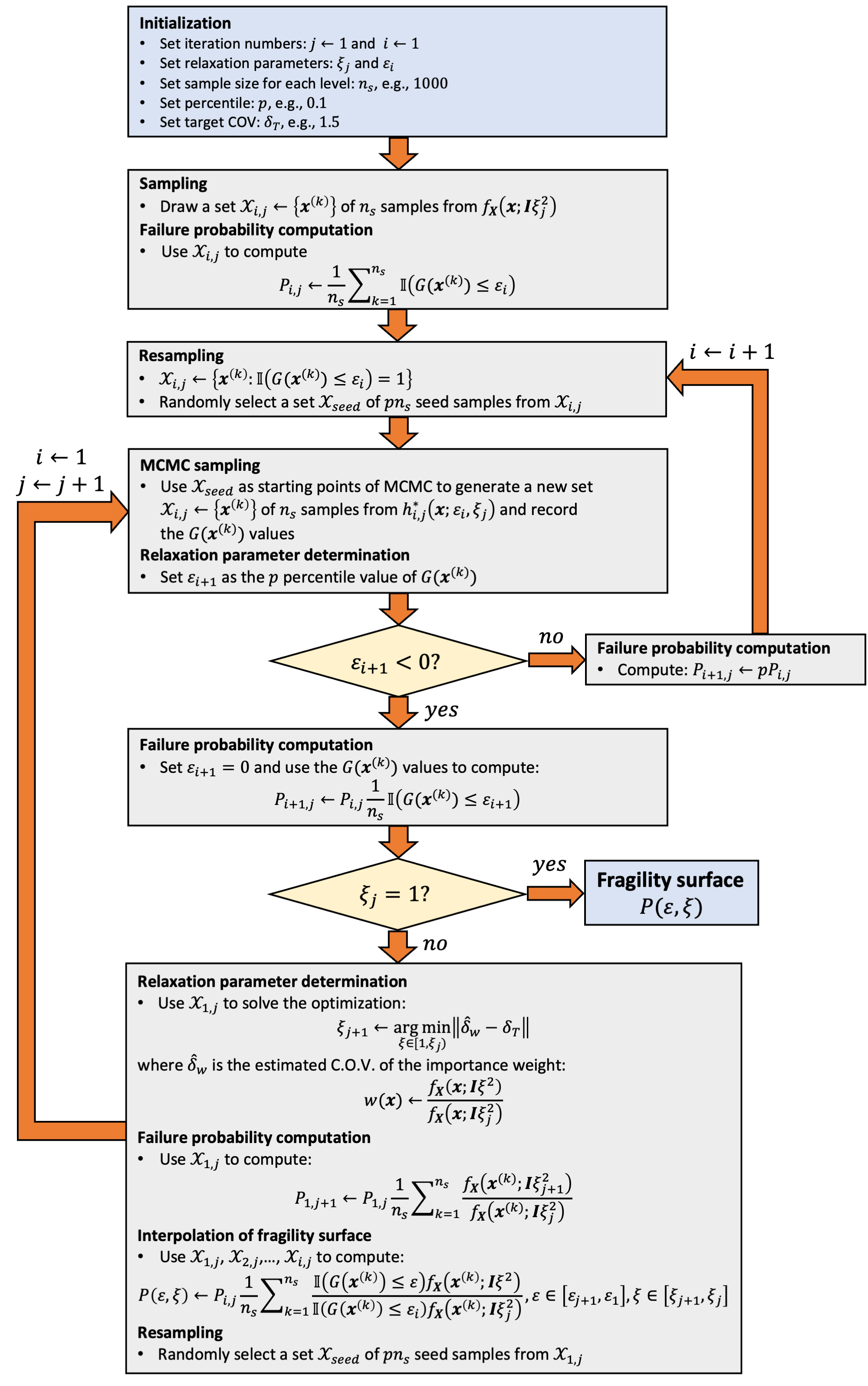}
			\caption{\textbf{Implementation details of IS-I for fragility surface estimation}.}
			\label{Fig:ISone}
		\end{figure}

		\begin{figure}[H]
			\centering
			\includegraphics[scale=0.65]{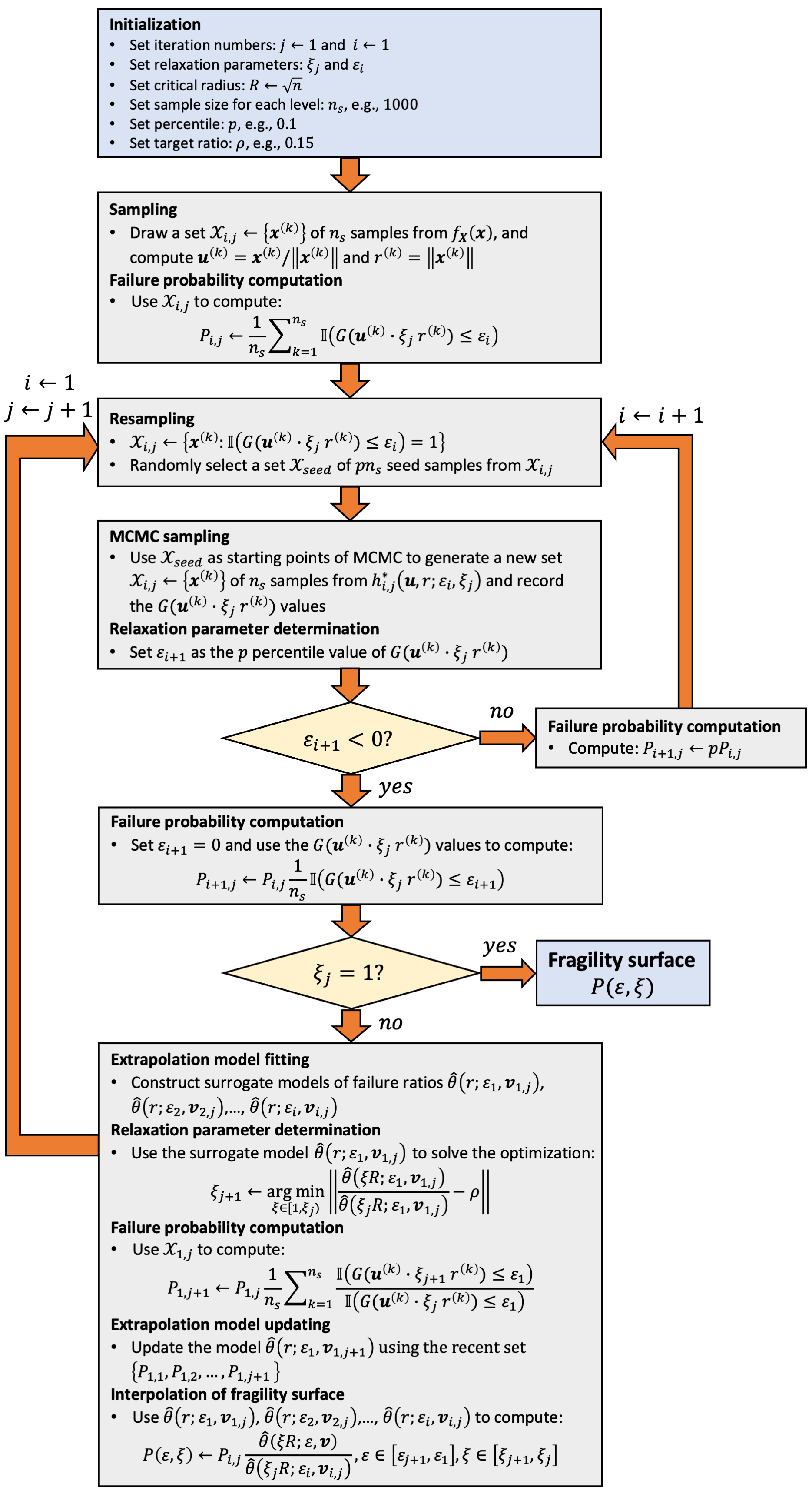}
			\caption{\textbf{Implementation details of IS-II for fragility surface estimation}.}
			\label{Fig:IStwo}
		\end{figure}
\end{document}